\documentclass[10pt]{bmc_article}    

\usepackage[english]{babel}

\usepackage{cite} 
\usepackage{url}  
\usepackage{ifthen}  
\usepackage{multicol}   
\usepackage[utf8]{inputenc} 
\usepackage{amssymb}
\usepackage{xcolor}
\usepackage{subfigure}
\usepackage{graphicx}
\usepackage{array}
\usepackage{multirow}
\usepackage{lmodern}
\usepackage{tabularx}
\usepackage[T1]{fontenc}
\usepackage{booktabs}
\usepackage{algorithmicx}
\usepackage{algorithm}
\usepackage{algpascal}
\usepackage{algc}
\usepackage{algcompatible}
\usepackage{algpseudocode}
\usepackage{pdflscape}
\usepackage{amsmath}
\usepackage{footnote}
\usepackage{rotating}
\usepackage{url}
\usepackage{subfigure}
\usepackage{caption}
\usepackage[colorlinks,urlcolor=blue,citecolor=blue]{hyperref}

\DeclareFontFamily{U}{mathx}{\hyphenchar\font45}
\DeclareFontShape{U}{mathx}{m}{n}{<-> mathx10}{}
\DeclareSymbolFont{mathx}{U}{mathx}{m}{n}
\DeclareMathAccent{\widebar}{0}{mathx}{"73}

\newboolean{publ}

\newenvironment{bmcformat}{\baselineskip20pt\sloppy\setboolean{publ}{false}}{\baselineskip20pt\sloppy}

\begin{document}
\begin{bmcformat}


\title{Automatic learning of pre-miRNAs from different species}
 

\author{Ivani de O. N. Lopes\correspondingauthor$^{1}$%
         \email{Ivani de O. N. Lopes\correspondingauthor - ivani.negrao@embrapa.br},
         Alexander Schliep$^2$%
         \email{Alexander Schliep - schliep@cs.rutgers.edu}%
       and
         André P. de L. F. de Carvalho$^3$%
         \email{André C. P. de L. F. de Carvalho - andre@icmc.usp.br}%
      }


\address{%
    \iid(1)Empresa Brasileira de Pesquisa Agropecuária, Embrapa Soja,\\
           Caixa Postal 231, Londrina-PR, CEP 86001-970, Brasil\\
    \iid(2)Department of Computer Science and BioMaPs Institute for Quantitative Biology \\
           Rutgers University\\
           110 Frelinghuysen Road, Piscataway, NJ, 08854, USA\\
    \iid(3)Instituto de Ciências Matemáticas e de Computação\\
           Avenida Trabalhador são-carlense, 400 - Centro, São Carlos - SP, Brasil
}%

\maketitle


\begin{abstract} 
        
\textbf{Background}: Discovery of microRNAs (miRNAs) relies on predictive models for characteristic features from miRNA precursors (pre-miRNAs). The short length of miRNA genes and the lack of pronounced sequence features complicate this task. To accommodate the peculiarities of plant and animal miRNAs systems, tools for both systems have evolved differently. However, these tools are biased towards the species for which they were primarily developed and, consequently, their predictive performance on data sets from other species of the same kingdom might be lower. While these biases are intrinsic to the species, the characterization of their occurrence can lead to computational approaches able to diminish their negative effect on the accuracy of \mbox{pre-miRNAs} predictive models. Here, we investigate in this study how 45 predictive models induced for data sets from 45 species, distributed in eight \mbox{subphyla/classes}, perform when applied to a species different from the species used in its induction. 

\textbf{Results}: Our computational experiments show that the separability of \mbox{pre-miRNAs} and pseudo \mbox{pre-miRNAs} \mbox{instances} is species-dependent and no feature set performs well for all species, even within the same \mbox{subphylum/class}. Mitigating this species dependency, we show that an ensemble of classifiers reduced the classification errors for all 45 species. As the ensemble members were obtained using meaningful, and yet computationally viable feature sets, the ensembles also have a lower computational cost than individual classifiers that rely on energy stability parameters, which are of prohibitive computational cost in large scale applications.

\textbf{Conclusion}: In this study, the combination of multiple pre-miRNAs feature sets and multiple learning biases enhanced the predictive accuracy of pre-miRNAs classifiers of 45 species. This is certainly a promising approach to be incorporated in miRNA discovery tools towards more accurate and less species-dependent tools.
\end{abstract}

The material to reproduce the results from this paper can be downloaded from \href{http://bioinformatics.rutgers.edu/Static/Software/multispecies.tar.gz}{http://bioinformatics.rutgers.edu/Static/Software/multispecies.tar.gz}.

\section*{Background}
\mbox{\textbf{MicroRNAs}} (miRNAs) constitute one of the most widely-studied class of endogenous small (approx. 22 nucleotides) non-coding RNAs genes, due to their regulatory role in pos-transcription gene regulation in animals, plants and fungi \cite{Lim2003b,Westholm2011}. The miRNAs biogenesis involves the participation of several enzymes, which depend on the origin (e.g. intergenic or intronic miRNAs) and on the kingdom of the species. However, all miRNAs are processed from long primary \mbox{miRNA} transcripts (pri-\mbox{miRNAs}), which are processed to hairpin-shaped intermediates (\mbox{pre-\mbox{miRNAs}}) and, subsequently, to the double strand RNA \mbox{miRNA}:\mbox{miRNA}* and a terminal loop. The \mbox{miRNA}* strand is the reverse complement of the functional \mbox{miRNA}, which usually degrades after being unwind by the action of specific enzymes. In the cytoplasm of animal and plant cells, the mature \mbox{miRNA} enters in the RNA-induced silencing complex (RISC) to silence target messenger RNAs (tmRNAs) by partial or near-perfect antisense complementarity. Partial antisense complementarity inhibits the translation of tmRNAs, whereas the later causes the degradation of tmRNAs. Reviews on biogenesis, diversification and evolution of miRNAs can be obtained at \cite{Bartel2004,Westholm2011,Berezikov2011}.

\textbf{ RNAseq methods, followed by computational analysis}, became the \textit{de facto} approach for miRNA discovery \cite{Berezikov2011}. These methods, also called deep sequencing of the transcriptome, can reveal the identities of most RNA species inside a cell, providing tens to hundreds of millions of sequence reads \cite{Chu2012}. These reads provide both the sequence and the frequency of RNA molecules present in a cell. When applied to detect \mbox{miRNAs}, the RNA material is isolated through a procedure of size selection, such that only small reads (approx. 25 nt long) are sequenced \cite{Chu2012}. The computational challenge consists in distinguishing \mbox{miRNAs} from other small RNA (sRNA) types and degradation products \cite{Friedlander2008,Berezikov2011}.

\textbf{The challenge of building a multi-species miRNA prediction tool} is reflected in the wide range of sensitivities estimated for eight deep sequencing miRNA prediction tools, when they were applied to data sets from \textit{H.~sapiens}, \textit{G.~Gallus} and \textit{C.~elegans} \cite{Li2012}. The sensitivity ranges varied between 24\% and 38\%. For example, the sensitivity of the tool with the highest average sensitivity (68\%) varied between 55\% (\textit{H.~sapiens}) and 78\% (\textit{G.~Gallus}) and the sensitivity of the tool with lowest average sensitivity (15\%) varied between 0\% (\textit{H.~sapiens}) and 25\% (\textit{C.~elegans}). The species bias is also present in the analysis performed with miRDeep2 \cite{Friedlander2012}, a newer version of miRDeep \cite{Friedlander2008}, which incorporated additional features to increase the detection of known and novel miRNAs in all animal major clades. Even though the average sensitivity of miRDeep2 (80\%) has clearly increased, compared to its first version, it still varies depending on the species from 71\% (Sea squirt) to 90\% (Anemone). In order to identify the source of these variabilities, it is imperative to explore how the main factors involved in the development of such computational tools vary throughout species.

\textbf{As miRNAs are processed from hairpin regions}, computational tools developed to predict miRNAs from RNA-seq libraries include at least four steps: pre-processing; read mapping to a reference genome; detection of energetically stable hairpins in the genomic region surrounding the mapped read and; detection of miRNAs \mbox{biogenesis `signature'}. The latter is derived from the abundance and from the distribution of the reads across the hairpin and is fundamental to reduce false detections, since the hairpin shape structure is a necessary but not sufficient condition to process miRNA. Three criteria have been used as evidence of miRNAs biogenesis: a) the frequency of the mature strand is higher than the frequencies of the corresponding star and loop strands; b) the positions of the Drosha and Dicer cleavage sites in the 5' ends of the putative miRNA and miRNA* are nearly uniform and; c) the putative miRNA and miRNA* sequences align in the hairpin keeping approximately 2 nt overhang in the 3' end \cite{Berezikov2011}. Nevertheless, the hairpin analysis is possibly the most critic step affecting negatively the sensitivity of the tools, since the biogenesis signature analysis is performed either after the selection of the energetically most favorable hairpin containing the mapped read stack (e.g. as in miRanalyzer \cite{Hackenberg2011}) or simultaneously, where the distribution of the reads in the putative hairpin and hairpin features are considered (as in miRDeep2 \cite{Friedlander2012}). 

\textbf{The hairpin analysis} has been performed mostly through machine learning based predictive models. To obtain these models, a feature set (feature vectors) describing sequence and/or structural aspects of pre-miRNAs sequences (+) and hairpin like (-) sequences is extracted to create a training data set, which is subsequently fed to a machine learning algorithm. An investigation on human pre-miRNAs classifiers indicated that the feature set, instead of the learning algorithm, had the major effect in the classification accuracy of the induced models \cite{Lopes2014}. However, the relevance of those features for the correct classification of pre-miRNAs from other species remained an open question. 

\textbf{Since miRNA systems in plants and animals differ substantially} \cite{Berezikov2011}, computational tools for plant and animal miRNAs discovery have been developed separately (eg. \cite{Hackenberg2011,Yang2011}). However, in practice, even instances of species from the same kingdom apparently diverge substantially regarding their intrinsic and extrinsic features. Therefore, in order to develop miRNA discovery tools robust to species-specific differences, a first step is to determine if a unique feature set can capture the diversification of pre-miRNAs throughout species. Moreover, it is important to establish the boundaries of the applicability of cross-species miRNAs predictive models, since the relevance of any tool depends on its ability to detect the miRNAs present in the data set under analysis. Another important aspect to be considered is the computational cost of extracting a feature set, since this cost can be prohibitive for some distinct pre-miRNAs features (e.g. energy stability parameters) it they are to be computed for millions of hairpins. These issues were addressed in this study, considering eight feature sets investigated in \cite{Lopes2014}, three learning algorithms and 45 species representing eight subphyla/classes.

\textbf{Our experimental results showed that the classification complexity of pre-miRNAs is species-dependent}, albeit some feature sets and learning algorithms were more likely to maximize the predictive accuracy of pre-miRNAs classifiers for most species (first subsection of the Results and Discussion section). To interpret this dependency, we analyzed how relevant the features extracted from instances of one species are for the classification of instances of other species (in the following subsections). This analysis indicated that pre-miRNAs classifiers restricted to predict instances of species from the same subphylum of the species used on its induction (training species), instead of the same kingdom, are more likely to achieve higher accuracies. Nevertheless, our results also showed that ensembles of classifiers using computationally inexpensive feature sets performed well even if the subphylum of the training species disagrees. The ensemble approach has the potential to extend the applicability of pre-miRNA predictive models to a broader number of species, while keeping the computational cost close to that of single classifiers.

\section*{Material and Methods}

\subsection*{Experimental design}
The analysis carried out in this study was based on the accuracy of classifiers obtained in two steps: (1) create pre-miRNA data sets and (2) induce and test classifiers for classification of pre-miRNAs. In the step (1), for each species, 30 sequences from each class were randomly sampled from the pre-processed positive and negative sets to compose the test sets. From the remaining sequences, 60 sequences from each class were randomly sampled to construct the training set. Afterwards, all features were extracted from each sequence. This first step was repeated 10 times. As these data sets were built by species, they are also referred as training and test species. In the step (2), instances from all test sets were classified by the classifiers obtained with the training data built in the step (1). The accuracy of these classifiers were analyzed under the two-way analysis of variance (anova) equations \ref{eq:anova1} and \ref{eq:anova2}.

The sizes of the training and test sets were, respectively, 2/3 and 1/3 of the smallest number of positive non-redundant sequences, shown in the Additional file 1. By fixing the sizes of training and test sets, we reduced the sources of random variations, i.e., variations that cannot be assigned to a main factor.  However, since our main goal was to study the effect of the training species ($S$) in the predictive accuracy of pre-miRNAs classifiers, we considered the effects of the classification algorithm and the feature set in a unique factor, represented here by $M$. Therefore, considering three algorithms and eight feature sets, the number of levels of the factor $M$ is 24 (or 3$\times$ 8). 

\subsubsection*{Anova 1: M $\times$ S}
The first analysis was performed to study the relationship $M \times S$ in order to identify the levels of $M$ that led to higher predictive accuracies for each species. For such, we considered the Equation \ref{eq:anova1}, where the accuracies were estimated considering the same training and test species.
\begin{align} \label{eq:anova1}
A_{ilk}&=\mu + M_{l} + S_{i} + MS_{li} + e_{lik},
\end{align}
such that: \\
$l = 1,...,24$ indexes the classifiers, \\
$i = 1,...,45$ indexes the species, \\
$k = 1,...,10$ indexes the repetition, \\
$A_{lik} = $ accuracy of the classifier $l$, obtained with the training species $i$ in the repetition $k$, \\
$\mu = $ overall mean accuracy,\\
$ M_{l} = $ effect of the classifier $l$,\\
$ S_{i} = $ effect of the species $i$, \\
$ MS_{li} = $ interaction between the effects of the classifier $l$ and the species $i$, and \\
$ e_{lik} = $ random error, or part of $A_{lik}$ that could not be assigned to the classifier $l$, the species $i$ and the repetition $k$; $e \sim N(0,\sigma^{2})$.

\subsubsection*{Anova 2: cross-species classifiers}
To investigate the suitability of instances from one species to build pre-miRNAs predictive models for other species, we fixed a classifier $l$, $l=1..24$, and varied the training and test species. The accuracies were analyzed according to Equation \ref{eq:anova2}:
\begin{align} \label{eq:anova2}
A_{lijk}&=\mu + M_{li} + T_{j} + MT_{lij} + e_{lijk},
\end{align}
such that: \\
$l$ indexes one out the 24 classifiers, \\
$i,j = 1,...,45$ indexes training and test species, \\
$k = 1,...,10$ indexes the repetition, \\
$A_{lijk} = $ accuracy of the classifier $l$, obtained with data from the species $i$, in predicting the classes of instances from the species $j$ in repetition $k$, \\
$\mu = $ overall mean accuracy,\\
$ M_{li} = $ effect of a species $i$, \\
$ T_{j} = $ effect of the species $j$, \\
$ MT_{lij} = $ effect of the interactions model species $l$ and test species $j$, and \\
$ e_{lijk} = $ random error, or part of $A_{lijk}$ that could not be assigned to the species $i$, the test species $j$ and the repetition $k$; $e \sim N(0,\sigma^{2})$.

\subsubsection*{Clustering algorithm}
The equations \ref{eq:anova1} and \ref{eq:anova2} are particularly useful to estimate the variance of random errors ($\sigma^{2}$). Once this variance is known, we can decide how typical the variances estimated from the controlled factors (e.g. $M$, $S$ and $MS$) are, compared to $\sigma^{2}$, using the $p$-value obtained from the $F$-test. In this work, significant $p$-values were lower or equal to 0.05 \mbox{($p\le \text{0.05}$)}. Since significant $p$-values of $F$-test on a factor only supports the inference that the at least two levels of that factor had different average effects, we applied a clustering algorithm due to Scott and Knott \cite{Scott1974} to identify the levels of each factor in equations  \ref{eq:anova1} and \ref{eq:anova2} that led to non-significantly different accuracies using the R package ScottKnott \cite{Jeli2013}. 

\subsection*{Data sets}
\subsubsection*{Positive sequences}
To construct positive data sets, we downloaded all pre-miRNAs from miRBase release 20. This release contains 24,521 miRNA loci from 206 species, processed to produce 30,424 mature miRNA products \cite{Kozomara2014}. However, only 65 species had at least 100 pre-miRNAs. From these 65 species, 48 had at least 90 non-redundant sequences (see criterion in the pre-processing subsection). Based on the availability of sequences that could be used to generate negative examples, positive sequences from only 45 species were considered. The identification of these species per \mbox{phylum/division}, \mbox{subphylum/class}, the acronyms used in their \mbox{identification}, the amount of available and non-redundant \mbox{pre-miRNAs}, the mean and the standard deviation of their sequence length are shown in Table \ref{tab:characteristics}, Additional file 1. 

\subsubsection*{Negative sequences}
 Negative data sets were constructed from a pool of 1,000 pseudo \mbox{hairpins} per species. These pseudo hairpins were excised from Protein Coding Sequences (CDS) or pseudo gene sequences, downloaded from the repositories Metazome v3.0, Phytozome v9.0 or NCBI, as detailed in the Additional file 2. The excision points were randomly  chosen in the interval $[0,L - l_{pse}-100]$, where $L$ was the sequence length of the CDS or pseudo gene and $l_{pse}$ was the length of the excised sequence. The number of pseudo hairpins of length $l_{pse}$ were determined in accordance with the length distribution of the available pre-miRNAs from each species. Afterwards, the excised sequence was evaluated for the resemblance with real pre-miRNAs. Sequences that passed the criteria described in the items \ref{item:r1} to \ref{item:r4} below were stored as pseudo hairpins, and those that failed any of these criteria were discarded. These criteria were:
\begin{enumerate} 
\item fold-back structure (FB); \label{item:r1}
\item $bp \ge \text{18}$, $bp$ = base pairing;  \label{item:r2}
\item $Q_{seq}\ge \text{0.9}$, $Q_{seq}$ = sequence entropy;  \label{item:r3}
\item Minimum Free Energy of folding ($MFE$) rules: \label{item:r4} \\
 $MFE_{l_{pse}} \le \text{-10.0} $, if $l_{pse} < \text{70}$  \\
 $MFE_{l_{pse}} \le \text{-18.0} $, if $\text{70} < l_{pse} \le \text{100} $  \\
 $MFE_{l_{pse}} \le \text{-25.0} $, if $l_{pse} > \text{100}$. 
\end{enumerate} 
$Q_{seq}$ was used to filter out meaningless sequences, since genomic sequences are usually contiguously padded with "N" characters and the $MFE$ rules were derived to accommodated the correlation between $MFE$ and $L$. 

\subsubsection*{Pre-processing}\label{subsubsec:preprocess}
Genes in a miRNA family can have sequence identity of 65\% or higher  \cite{Kamanu2013}. Since the number of miRNA families is relatively small compared to the number of positive examples available, redundancy removal is an important pre-processing procedure to avoid overfitted predictive models. We used \mbox{dnaclust} \cite{Ghodsi2011} to remove redundant sequences, prior to the sampling of examples to compose training and test sets. With dnaclust, sequences in positive sets of each species were clustered such that the similarity between sequences within a cluster were at least 80\%. Afterwards, one sequence from each cluster was randomly sampled to construct the positive non-redundant sets. The same pre-processing procedure was applied to the sets of negative sequences. As detailed in Table \ref{tab:charpseudo} in the Additional file 2, 15 or less sequences were removed from the 35 out of 45 sequence sets. The relatively lower number of redundant pseudo hairpins in those sets, compared to pre-miRNAs, is due to the random choice of the starting position of the pseudo hairpin excision. However, at least 35 redundant pseudo hairpins were removed from the other 10 sequence sets. 

\subsection*{Feature sets} \label{subsec:featuresets}
The eight features sets primarily studied in this investigation were extensively evaluated on human sets by Lopes et al. \cite{Lopes2014}. Here, these feature sets are referred by the same notation (FS$_i$, $i\in \{1,..,7\}$ and SELECT). These feature sets contain most of the features used in computational pipelines for pre-miRNA discovery. References of computational pipelines that used these feature sets and their composition can be seen in Table 1. This table also shows two important aspects of these feature sets: feature diversity and feature sets overlapping. For example, FS$_1$, FS$_2$, FS$_7$ and SELECT have 13 overlapping features, from which five are also in  FS$_3$. Features in these sets are measures of different characteristics of the sequences, whereas the features in FS$_4$, FS$_5$ and FS$_6$ are mostly sequence-structure patterns.

\subsection*{Learning algorithms}
The learning algorithms used in this work were Support Vector Machines (SVMs), Random Forest (RF) and J48. These algorithms have different learning biases, which is important for the present work, since learning biases may favor a feature set over others. SVMs and RFs are the algorithms most used for pre-miRNA classification and J48 was chosen because of its simplicity and interpretability.

J48 implements the well known C4.5 algorithm \cite{Quinlan1993}. As one of the most popular algorithm based on the divide-and-conquer paradigm, C4.5 recursively divides the training set into two or more smaller subsets, in order to maximize the information entropy. The J48 implementation builds pruned or unpruned decision trees from a set of labeled training data. We used RWeka \cite{Rweka2009}, an R interface of Weka \cite{Weka2005}, with the default parameter values. RWeka induces pruned decision trees from a data set.

To train SVMs, we used a Python interface for the library LIBSVM 3.12 \cite{CC01a}. This interface implements the C-SVM algorithm using the RBF kernel. The kernel parameters $\gamma$ and $C$ were tuned by 5-fold cross validation (CV) over the grid ($C; \gamma$) = ( 2$^{-5}$,2$^{-3}$,...,2$^{15}$; 2$^{-15}$,2$^{-13}$,...,2$^{3} $). The pair ($C; \gamma$) that led to the highest CV predictive accuracy in the training subsets was used to train the SVMs using the whole training set. The resulting classifier was applied to classify the instances from the corresponding test set.

RF ensembles were induced over the grid (30, 40, 50, 60, 70, 80, 90, 100, 150, 250, 350, 450)$\times$[ (0.5, 0.75, 1, 1.25, 1.5)*$\sqrt{d}$ ], representing respectively the number of trees and the number of features. The value $\sqrt{d}$ is the default number of features tried in each node split, where $d$ is the dimension of the feature space or the number of features in the feature set. We chose the ensemble with the lowest generalization error over the grid, according to the training set, and applied it to classify the instances of the corresponding test set. The ensembles were obtained using the \textit{randomForest} R package \cite{Liaw2002} in \textit{in house} R pipeline.

\subsubsection*{Ensembles and other feature sets}
Besides the predictive accuracy, the applicability of any pre-miRNA classifier to larger data sets may be \mbox{limited} by the computational time necessary to compute the feature set representation of each \mbox{pre-miRNA} candidate. To increase the predictive accuracy while keeping the computational cost \mbox{under} feasible limits, subsets of the existing features sets, removing features computed from shuffled sequences, were employed to construct ensemble of classifiers. These subsets were named Ss1 and Ss7, such that: \mbox{$\text{Ss1} = \text{FS}_{1} - \{zG,zP,zQ,zD,zF\}$} and $\text{Ss7}=\{orf,\%LCRs,loops,A_{(((},C_{(((},G_{(((},U_{(((}\}$. Ss1 features \mbox{measure} the largest variety of \mbox{pre-miRNA} characteristics, whereas Ss7 combine features widely used in \mbox{pre-miRNA} classification $(A_{(((},C_{(((},G_{(((},U_{(((}$) with three features introduced in \mbox{pre-miRNA} classification in \cite{Gudy2013}. The first subset was evaluated individually, and combined with the latter (Hyb$_{17}=Ss7 \cup Ss1$). The subset Ss7 was also combined with the feature sets FS$_{3}$  ($\text{Hyb}_{37}=\text{FS}_{3}\cup\text{Ss7}$) and SELECT (\mbox{$\text{Hyb}_{S7}=\text{Ss7}\cup\text{SELECT}$)}. The prefix Hyb is used to represent these `hybrid' feature sets.

An ensemble of classifiers combine the prediction of a set of individual classifiers. The ensembles used in this study are described in Table 2, along with all other classifiers investigated. The computational time for the extraction of the feature sets used in the ensembles are close to the time spent to extract the feature set SELECT and presented in \cite{Lopes2014}. As shown in this table,  the final prediction of the ensembles were defined by the majority vote (ensemble Emv) and weighted vote (ensemble Ewv). Each ensemble, therefore, combines the class prediction, class vote, from each one of its classifiers. In the first approach, the class predicted by the majority of the classifiers is the ensemble class prediction.  In the weighted approach, the vote of each classifiers was weighted by its predictive accuracy in the training set. Ties were resolved by random choice.

\section*{Results and discussions}

\subsection*{Predictive accuracy of pre-miRNA classifiers by species}
As the \textit{F}-test on the effect of $MS$ in Equation \ref{eq:anova1} was highly significant ($p<\text{0.001}$), the effect of the simple factor $M$ was studied within fixed levels of $S$ (M/S$_j$, $j=\text{1},...,\text{45}$), and vice-verse (S/M$_l$, $l=\text{1},...,\text{24}$). The analysis of M/S$_j$, $j=\text{1},...,\text{45}$, is summarized in Figure 1 and Table 3. The green bars in Figure 1 indicate the pre-miRNA classifiers whose accuracy is within the cluster of maximal accuracies $C_1$. As indicated in Figure 1, SVMs and RFs obtained using the feature sets FS$_{3}$, FS$_{6}$, FS$_{7}$ and SELECT achieved accuracies within $C_1$ for most species. These results agree with the results reported in \cite{Lopes2014}, which used larger training and test sets of human instances. 

Figure 1 indicates only the algorithms and feature set combinations more likely to produce pre-miRNAs classifiers of maximal accuracy, but the maximal depends on the species, as it can be observed in Table 3. According to this table, the mean accuracy in $C_1$ varied from 86\% (cin) to 96\% (ssc). As the clusters were obtained for each species using the estimated accuracies of the same 24 classifiers and the number of clusters varied from two (bfl, dme, hsa, ath, lus, mdm, ptc, osa, zma) to five (gga), Table 3 indicates that either the instances from some species are easier to classify than instances from other species, or that pre-miRNAs of different species carry specific features that identify related characteristics. In both cases, these results indicate that the incorporation of intrinsic characteristics of the species could improve the accuracy of \mbox{pre-miRNAs} predictive models in the classification of sequences from different species.

Table 4 presents the results of the analyzes of S/M$_l$, $l=\text{1},...,\text{24}$. Similar to what was observed in the analyzes of M/S$_j$, $j=\text{1},...,\text{45}$, the number of clusters and the corresponding centers depended on the levels of $M$. However, the number of clusters and the accuracy intervals (Range columns) in both tables show that the effect of $S$ in the accuracy of pre-miRNA classifiers is broaden than the effect of $M$.  For example, the number of clusters in Table 4 varied from two to six and the ranges varied from 14\% (FS$_7$-RFs) to 41\% (FS$_1$-J48). Moreover, although the average accuracies estimated from 17 out of 24 pre-miRNA classifiers were above 95\% for some species (column $c_1$), the average accuracies of the same level $M_i$ for other species were as low as 57\%. In fact, no $M_l$, $l=1,...,24$ led to classifiers of accuracies within $c_1$ for all species, supporting again the conjecture that the learning complexity of pre-miRNAs is species-dependent. 

In the next subsection, we discuss how representative the instances from the 45 species considered in this work are for the induction of classifiers able to predict the classes of each other's instances, given a classification algorithm and a feature set. In addition, we discuss the occurrence of species-specific features and their effect in the predictive accuracy of cross-species pre-miRNAs classifiers.

\subsection*{Cross-species pre-miRNAs classifiers: M$_{l}$ $\times$ T}
Given a learning algorithm and a feature set, the relevance of the instances of a species $i$ (training species) in the prediction of instances from a species $j$ (test species), $i\ne j$, can be inferred from the effects of the factors in Equation \ref{eq:anova2}. Since the \textit{F}-test on the interaction $M_{l}T$ was significant ($p\le \text{0.05}$), the factor $M_l$ was analyzed within each level of the factor $T$ ($M_l/T_j, j=\text{1},...,\text{45}$), and vise-verse ($T/M_{li}, i=\text{1},...,\text{45}$). The results of the analyzes of $M_{l}/T_j, j=\text{1},...,\text{45}$ indicate the training species that resulted in pre-miRNA classifiers of higher accuracies (c$_1$) for each test species. 
From the results of the analyzes of $T/M_{li}, i=\text{1},...,\text{45}$, we discussed the learning complexity of pre-miRNAs from the 45 species. 

\subsubsection*{Choosing the training species -  M$_{l}$/T}
By clustering the average accuracies $\widebar{A}_{lij\cdot}$, within $j$, $i, j=\text{1,...,45}$, we identified the training species $i$ that led to accuracies within c$_1$ for each test species $j$. Figure 2 shows these cases in green (c$_1$) and red (c$_2$,...,c$_6$), where $i$ is shown in the $Y$-axis and $j$ in the $X$-axis. The results for the other 20 models were similar. As the black frames enclose species from the same subphylum/class and within each frame the green pixels are more numerous than the red ones, we conclude that a pre-miRNAs classifiers was more likely to achieve predictive accuracies within c$_1$ when the species $i$ and $j$ were from the same subphylum/class. In particular, all means $\widebar{A}_{lij\cdot}$ were in c$_1$ when $i=j$ (diagonal), indicating that species-specific classifiers is a good approach to improve the predictive accuracy of pre-miRNAs predictive models. 

Figure 2 also shows that instances from some species were systematically harder to classify than instances from other species, which can be inferred through the number of red pixels per column. Among them, instances from bmo were typically harder to classify than instances from other species. The columns showing the clusters associated with different training species in the classification of instances from B. mori (bmo) and L. usitatissimum (lus) illustrate these cases. Particularly, the average of the clusters obtained from SVMs\_SELECT classifiers generated with instances of all species in predicting the classes of bmo instances were 80\% (c$_1$), 70\% (c$_2$) and 65\% (c$_3$), whereas the corresponding measures for lus were 98\% (c$_1$), 93\% (c$_2$), 89\% (c$_3$), 80\% (c$_4$) and  65\% (c$_5$). 

Although the phylogenetic proximity of training and test species is fundamental to obtain pre-miRNAs classifiers of higher accuracies,  the learning biases of the classification algorithm may increase or decrease the relevance of the subphylum/class membership, as Figure 2 shows. In this figure, SVMs were more sensitive to the phylogenetic proximity of training and test species. An interpretation for this pattern is provided in the subsection Feature importance.

\subsubsection*{Inferring learning complexity - T/M$_l$}
In these comparisons, we clustered the accuracies estimated from all test sets, fixing the training species and a level of $M$. These clusters are displayed in Figure 3, for four levels of $M$. In this figure, a row shows the test species ($X$-axis) assigned to the cluster $c_1$ (green) or to another cluster (orange), when its instances were classified using a training species $i$ ($Y$-axis). The highest quantities of green pixels clearly associated with the Angiosperm test species suggest that instances from Angiosperm test species were easier to classify than instances from other test species, particularly vertebrates.

Although this pattern was consistent in all 24 level of $M$, we also looked into the learning complexity by analyzing the importance of the 85 unique features in the classification of instances from all species. The idea was to indirectly compare the similarities between the instances from different species, using a feature importance measure obtained during the induction of RF classifiers. These results are discussed next. 

\subsection*{Feature importance}
Given a feature set, the importance of each feature for the correct classification of the test set instances can be estimated by a feature importance measure, which in this work was taken from the RF results. The rationale of investigating the relevance of the RNA features used in this work for the correct classification of pre-miRNAs of different species is to infer, at least indirectly, if the phylogenetic proximity of these species is a valid criterion to choose a feature set. 

The feature importance measure ($FI$) used in this study estimates the increase of misclassified OOB (Out-Of-Bag) instances when that feature is permuted in the training vectors. Since that measure is an absolute value, to allow its comparison for different classifiers induced with instances of different species, its values were re-scaled to the interval [0, 1] by the formula $RFI=(FI-FI_{min})/(FI_{max}-FI_{min})$. The maximum ($FI_{max}$) and minimum ($FI_{min}$) $FI$ values were obtained from the subset of features used in the induction of each pre-miRNA classifier. We estimated the $RFI$ values for each of the 85 unique features considered in this work feature, when they were simultaneously fed to the RF algorithm to induce pre-miRNA classifiers for each of the 45 species. These estimates were discussed based on two criteria: the pairwise Pearson correlation coefficient between species and the distributions of the $RFI$ for the 45 species.

\subsubsection*{Pearson correlation coefficients of RFIs between species}
Figure 7 shows the pairwise Pearson correlation coefficients of $RFI$ for all pairs of species. These correlations are in the interval [0, 1], where the black pixels indicate zero correlation and the white pixels indicate correlation one. Therefore, white or light gray pixels represent the cases where the pre-miRNAs of the two corresponding species shared most of the features. As the red frames indicate, these cases are more likely if the two species are from the same subphylum/class. However, there are many exceptions within and outside the subphylum/class umbrella. For example, with few exceptions (e.g. ame, bmo and bta), the features that are important for the correct classification of instances from the species bfl, cin, cbr, cel and aae, were also important for the correct classification of instances from other species. Differently, the difficulty in establishing a general rule on the association between phylogenetic proximity and feature conservation using the $RFI$ criteria can be observed by the majority of dark pixels associated with Hexapoda species. This exceptions and the features with the highest $RFI$ are presented next.

\subsubsection*{RFI distributions}
The $RFI$ distributions are shown in Table 5, omitting the cases where $RFI\le$0.1 for all species. According to this table, only 40\% of the features met this criterion. Among them, $p$ and $MFEI_{1}$ obtained $RFI$ larger than 0.6 for 89\% ($p$) and 94\% ($MFEI_{1}$) of the species, whereas the $RFI$ distributions of the other features were closer to a right-tailed distribution. In fact, the $RFI$ estimates of 75 out of 85 of features were lower than 0.3 for 80\% of the species. These small amount of highly relevant features helps to interpret the tendency of SVMs to reduce the predictive accuracy when the training and the test species were more distantly related, as those from Chordate and Angiosperm (Figure 2). Since SVMs use the full feature space and RFs use only subspaces of it, the classification by RFs may have been dominated by features that are more conserved throughout species. The interactions between the learning biases and the species is also analyzed through the classification errors of the three learning algorithms in the next subsection.

\subsection*{Classification error}
The classification errors of a particular instance by different classifiers can provide information on how typical that instance is, assuming that atypical instances or outliers are more likely to be misclassified by most classifiers. Moreover, the classification errors estimated from test sets of instances from different species by multiple classifiers is also informative of the separability of classes, in the instance space of each species. To facilitate the notation, the errors $e_1, e_2,.., e_7$ are defined as exclusive classification errors of SVM ($e_1$), RF ($e_2$), J48 ($e_3$), SVM and RF ($e_4$), SVM and J48 ($e_5$), RF and J48 ($e_6$) and SVM and RF and J48 ($e_7$). Since $e_{1},...,e_{7}$ are exclusive errors, they sum one or 100\%, symbolically: $\sum_{i=1}^{7} e_{i}=1$ or $\sum_{i=1}^{7} e_{i}=100\%$. These errors are shown in Figure 4, for FS$_{1}$, FS$_{6}$ and \mbox{SELECT}.

As can be observed in Figure 4, the error distributions were strongly dependent on the species, which shows in another way the classification biases associated with species sequence data. For example, Figure 4(a) shows that $e_1$ was zero for 15 species (cbr, tca, aca, gga, eca, ggo, ptr, cgr, ppt, aly, ath, mdm, ptc, osa, zma). Nevertheless, this same figure also shows $e_1$ of up to 80\% for other species (e.g., bfl, cin, ame, mtr, stu, sbi). In these cases, and others where the exclusive error of a classifier induced by one of the three algorithms is higher than the errors achieved simultaneously by at least two classifiers induced by different algorithms, the separability of the classes is a matter of choosing an algorithm with the appropriate learning bias. On the other hand, the cases where $e7>\text{50}\%$ (e.g. mdm) could be better described by other feature spaces or by a combination of subspaces.

To summarize, the classification errors in each feature space, the errors $e_{1},...,e_{7}$, were summed up for the 45 species and represented in Venn diagrams. Figure 6 shows the cases FS$_{1}$, FS$_{6}$, FS$_{7}$ and \mbox{SELECT}. The interaction between learning algorithm and feature set, indicated by the significant variation of the areas of the circles, between feature sets, is the most noticeable pattern in this figure. For example, classification models induced by J48 tended to achieve higher exclusive error rates ($e_{3}$) in higher dimensional feature spaces. Moreover, $e_{7}$, the proportion of instances misclassified simultaneously by classifiers induced by the three algorithms varied by 25\% between 3.2\% and 6.7\% ($\text{3.7}\% \le e_{7} \le \text{6.7}\%$). These two factor alone are sufficient to conjecture that  the combination of multiple hypotheses may lead to pre-miRNA classifiers of higher accuracies than a single hypothesis, for a larger number of species. To provide a preliminary insight on this conjecture, we carried out additional computational experiments, using ensemble approaches to combine multiple hypothesis to improve the predictive accuracy of pre-miRNA classifiers. These results from these experiments are presented and discussed in the next subsection.

\subsection*{Ensembles}
Figure 5 shows the comparisons between the 44 classifiers, as defined in Table 2. According to Figure 5, the ensembles Emv24, Ewv24, Emv8-RF, Emv8-SVMs, Ewv8-RF, Ewv8-SVMs, Ewv24 and the classifiers obtained with the new feature sets presented better predictive accuracies than the 24 previously discussed, for many species, although none of the them achieved predictive accuracies within C$_{1}$ for all 45 species. Moreover, it is important to remind that these ensembles and the new feature sets do not include features extracted from shuffled sequences. Figure 5 also shows that the simple combination of different hypotheses can increase the predictive accuracy, even using the algorithm J48, which typically led to equal or lower classification accuracies than RFs and SVMs.

Based on the results shown in figures 5 and  7, and in Table 5, we can state that it is unlike that a unique learning algorithm and a unique set of features is able to produce the best pre-miRNA predictive model for all species. In fact, the experimental results obtained in this study suggested that the learning of good predictive models for pre-miRNAs classification depends on the learning complexity inherited of the problem and the peculiarities of the instances from different species. Since ensembles apparently provide an alternative and efficient approach to accommodate these peculiarities, an appropriate construction of hypothesis diversity (e.g. \cite{Ding2010}) may enhance the performance of miRNA discovery tools in the classification of pre-miRNAs of different species. 

\section{Conclusion}
The increase in sequencing capacity and the computational analysis of large amounts of sequencing data to detect miRNAs supported the recent advances in the discovery of novel miRNAs from over a hundred species. Albeit miRNA systems vary throughout species, miRNA discovery tools from the literature have not addressed the impact of these differences. As a consequence, the performance of these tools is usually reduced when data sets from species not used in their development are analyzed. Building species-specific miRNA discovery tools may not be always viable, for example for lack of training data. Since the detection of putative pre-miRNAs is an important step in the development of miRNA discovery tools, it is important to investigate how the peculiarities naturally occurring in pre-miRNAs between species relate to the learning bias of machine learning approaches. In this study, we presented the results of a systematic investigation on the automatic learning of pre-miRNAs of 45 species, using techniques traditionally employed by miRNA discovery tools from the literature. The results presented in this study not only showed the need to develop new approaches to handle the intrinsic characteristics of pre-miRNAs from different species, but we also indicated the way to go forward, using ensemble methods built with computationally efficient features.

\section{Competing interests}
The authors declare that they have no competing interests.

\ifthenelse{\boolean{publ}}{\end{multicols}{2}}{}


\bigskip

\section*{Author's contributions}
AS and AC conceived and supervised the study. IL assembled the data, implemented the scripts, ran the experiments and summarized the results. The three authors 
wrote and approved the final manuscript.

\section*{Acknowledgements}
  \ifthenelse{\boolean{publ}}{\small}{}
  We thank Empresa Brasileira de Pesquisa Agropecuária (Embrapa Soybean) for the continuum financial support to the first author.
 

\newpage
{\ifthenelse{\boolean{publ}}{\footnotesize}{\small}
 \bibliographystyle{bmc_article}  
  \bibliography{biblio} }     
  
 \section*{Figures}
 
  \subsection*{Figure 1 - Frequencies of species for who each classification model achieved accuracies in the clusters C$_{1}$-C$_{5}$. Mean$_{\text{C}_{1}}\ge .. \ge$Mean$_{\text{C}_{5}}$.}
\begin{figure} \centering
  \includegraphics[scale=0.55]{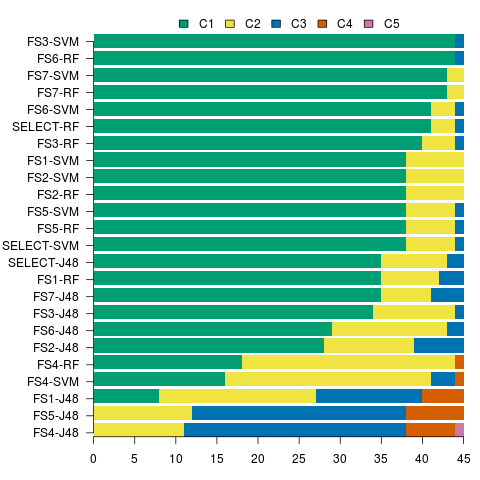}
     \label{fig:freq24}
\end{figure}

  \subsection*{Figure 2 - Accuracy cluster membership (columns) for cross-species pre-miRNAs classifiers. Green= c$_1$; red=other; $y$-axis=model species; $x$-axis=test species; black frames encloses species from the same subphylum/class.}
\label{fig:cross_species_test}
\begin{figure}
{\includegraphics[scale=0.525]{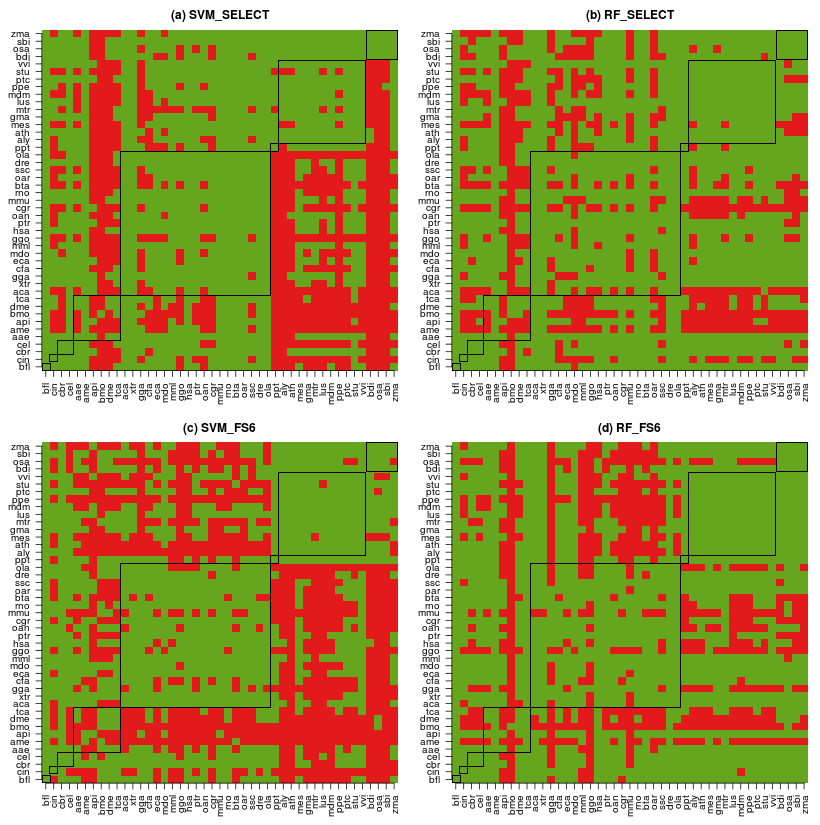}}
\label{fig:cross_species_test} 
\end{figure}

  \subsection*{Figure 3 - Accuracy cluster membership (rows) for cross-species pre-miRNAs classifiers. Green= c$_1$; red=other; $y$-axis=model species; $x$-axis=test species; black frames encloses species from the same subphylum/class.}
\begin{figure}
{\includegraphics[scale=0.525]{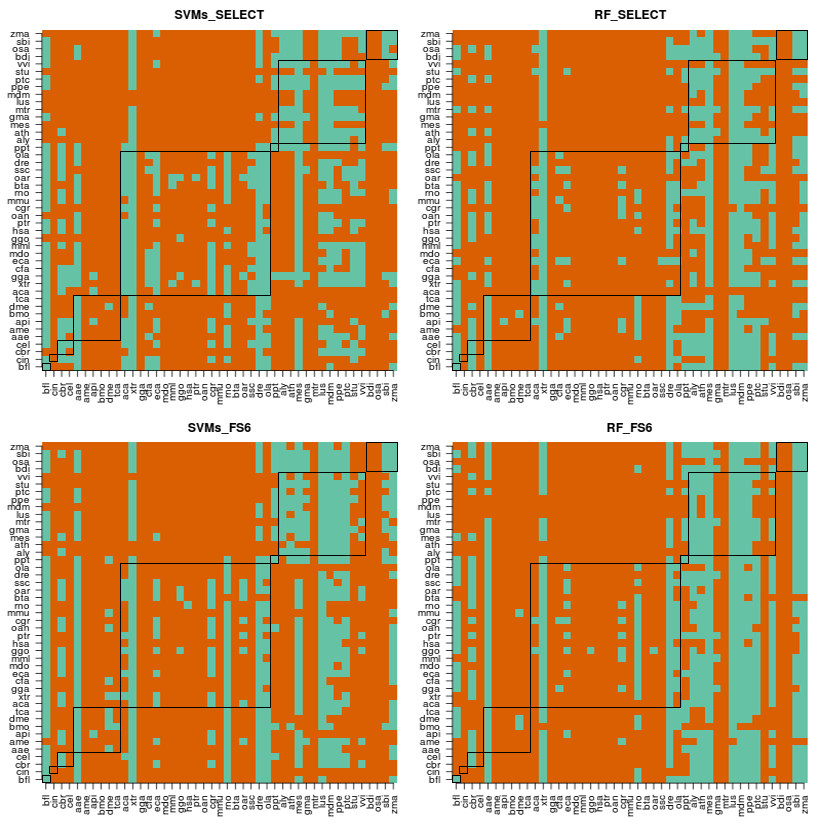}}
\label{fig:cross_species_tr} 
\end{figure}

\subsection*{Figure 4 - Distribution of classification errors per species. Exclusive errors by SVMs ($e_1$), RF ($e_2$), J48 ($e_3$), SVMs and RF  ($e_4$), SVMs and J48 ($e_5$), RF and J48 ($e_6$) and SVMs and RF and J48 ($e_7$).}

\begin{figure}
{\includegraphics[scale=0.475]{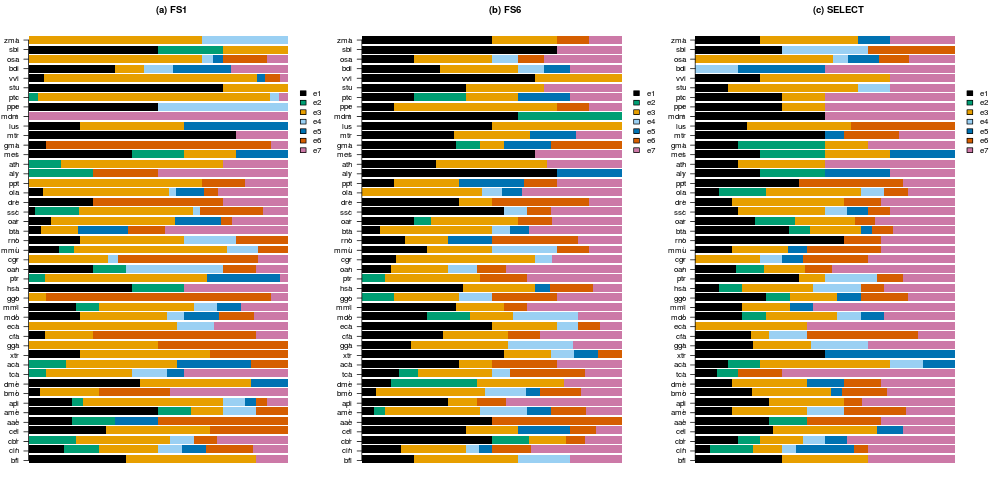}}
\label{fig:errorsprofile} 
\end{figure}

\subsection*{Figure 5. Distribution of the accuracies of 44 classifiers within the accuracy clusters. $\text{Mean}_{C_1} > ... > \text{Mean}_{C_5}$.}

\begin{figure}
  \centering
  \includegraphics[scale=0.5]{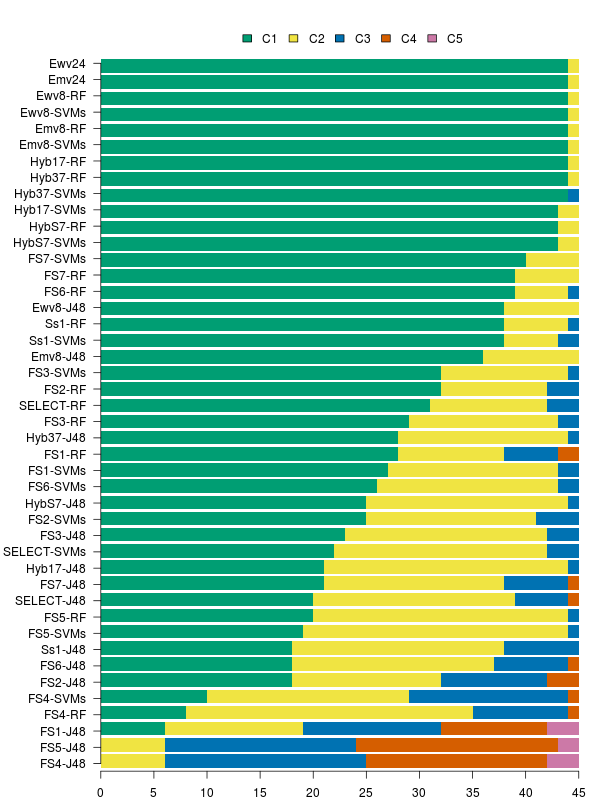}
  \label{fig:compareall}
\end{figure}

\subsection*{Figure 6 - Venn diagram of the classification errors of the classification algorithms, by feature set. Results were obtained from the classification of 27,000 = 45 (test species) $\times$10 (repetitions) $\times$60 (30+,30-).}
\begin{figure}
\includegraphics[scale=0.4]{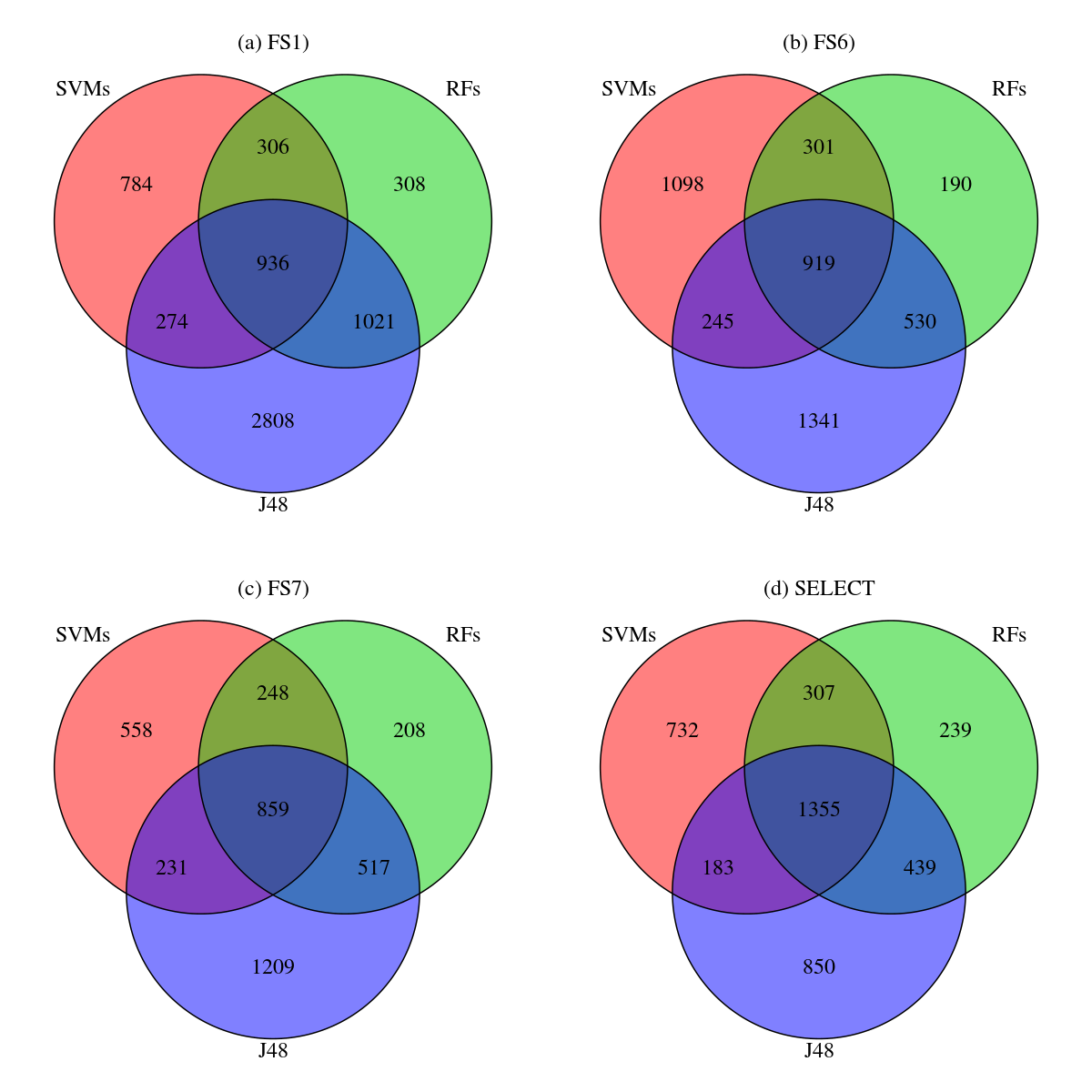}
\label{fig:venn}
\end{figure}

\subsection*{Figure 7 - Pairwise Pearson correlation coefficient of $RFI$ throughout species.}
\begin{figure}
 \includegraphics[scale=0.5]{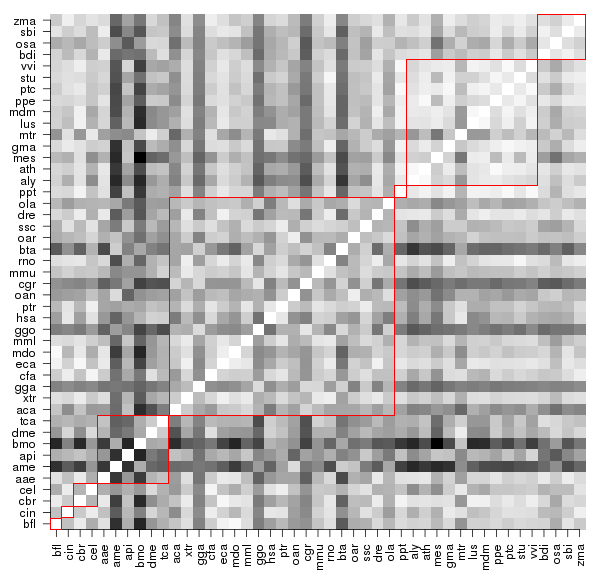}
\label{fig:frcorrelation}
\end{figure}







\section*{Tables}
   
 \subsection*{Table 1 - Feature set composition, dimension, literature reference and associated literature tool.}   
\begin{table}
  \small
  \label{tab:features}
  {\scriptsize
     \begin{tabularx}{\textwidth}{ X | c | c | c | c | c | c | c | c }  \noalign{\hrule height 1.0pt}   
        \multirow{2}{*}{FEATURE}  & \multicolumn{8}{c}{FEATURE SET} \\ \cline{2-9}
 & FS$_1$ & FS$_2$ & FS$_3$ & FS$_4$ &FS $_5$ & FS$_6$ & FS$_7$ & SELECT \\ \noalign{\hrule height 1.0pt}
         Di-nucleotide frequencies ($XY, X,Y \in \{A,C,U,G\}$) & x &   &   &   &   &    &     &\\ 
         $\%G+C$ & x & x &   &   &   &    & x  &\\ 
  Maximal length of the amino acid string without stop codons ($orf$) & & & & & & & x &\\ 
         Percentage of low complexity regions ($\%LCRs$) &    &   &   &   &   &    & x  &\\ 
         Triplets &    &   &   & x &   & x &     &\\ 
         Stacking triplets ($X_{(((}$, $X \in \{A,C,G,U\}$) &    &   &    &   &   &   & x  &\\ 
         Motifs ($ss-$substrings) &    &   &    &   & x &   &    &\\ 
         Minimum free energy of folding ($MFE$) &    &   &    &   &   & x &    &\\
         Randfold ($p$) &    &   &    &   &   & x &    &\\ 
         Normalized MFE ($dG$) & x  & x & x &   &   &   & x  & x \\
         MFE index 1 ($MFEI_{1}$) & x  & x & x &   &   &   & x  & x \\ 
         MFE index 2 ($MFEI_{2}$) & x  & x & x &   &   &   & x  & x \\ 
         MFE index 3 ($MFEI_{3}$) & x  & x &   &    &   &   & x  & x \\
         MFE index 4 ($MFEI_{4}$) &  x & x &   &    &   &   & x  &\\
         Normalized Ensemble Free Energy ($NEFE$) & x  & x &   &    &   &   & x  & x \\
         Normalized difference ($MFE-EFE$) ($Diff$) & x  & x &   &    &   &   & x  & x \\
         Frequency of the MFE structure ($Freq$) & x  &   &   &    &   &   &     &\\ 
         Normalized base-pairing propensity ($dP$) & x  &   & x &    &   &   &     &\\
         Normalized Shannon entropy ($dQ$) & x  & x & x &    &   &   & x  & x \\
         Structural diversity ($Diversity$) & x  & x &   &    &   &    & x  &\\
         Normalized base-pair distance ($dD$) &  x  &   & x &    &   &    &    &\\ 
         Average base pairs per stem (Avg\_Bp\_Stem) &  x  & x &   &    &   &    & x  &\\ 
         Normalized A-U pairs counts ($|A-U|/L$) &  x  & x &   &    &    &   & x  &\\ 
         Normalized G-C pairs counts ($|G-C|/L$) &  x  & x &   &    &    &   & x  & x \\ 
         Normalized G-U pairs counts ($|G-U|/L$) &  x  & x &   &    &    &   & x   & x \\ 
         Content of A-U pairs per stem ($\%(A-U)/stems$) &  x  & x &   &    &    &   &  x  & \\ 
         Content of G-C pairs per stem ($\%(G-C)/stems$) &  x  & x &   &    &    &   & x   &\\ 
         Content of G-U pairs per stem ($\%(G-U)/stems$) & x & x &  & & &  & x  & x \\ 
         Cumulative size of internal loops ($loops$) &     &   &    &    &    &    & x   &\\ 
         Structure entropy ($dS$) &  x  & x &   &    &    &    & x  & x \\ 
         Normalized structure entropy ($dS/L$) &  x  & x &   &    &    &    & x  & x \\
         Structure enthalpy ($dH$) &  x  &   &   &    &    &    &     &\\
         Normalized structure enthalpy ($dH/L$) &  x  &   &   &    &    &    &     &\\ 
         Melting energy of the structure &  x  &   &   &    &    &    &     &\\        
         Normalized melting energy of the structure &  x  &   &   &    &    &    &     & \\      
         Topological descriptor (dF) &  x  & x & x &   &    &    &  x  & x \\ 
         Normalized variants ($zG$, $zP$ and $zQ$) &  x  &    &   &   &    &    &      & \\ 
         Normalized variants ($zD$) &  x  & x &   &   &    &    &   x  &\\ 
         Normalized variants ($zF$) &  x  &    &   &   &    &    &       &  \\ \noalign{\hrule height 1.0pt} 
         DIMENSION &48  &21 & 7 &32& 1300 & 34 & 28 & 13 \\ \hline
         REFERENCE &\cite{Batuwita2009} & \cite{Batuwita2009} &\cite{Hsieh2010}&\cite{Xue2005}& \cite{Liu2012} & \cite{Jiang2007}& \cite{Gudy2013} & \cite{Lopes2014} \\  \noalign{\hrule height 1.0pt} 
  \end{tabularx}  } 
  \end{table}

\subsection*{Table 2 - Definition of all 44 classification models compared in this work, according to feature sets and learning algorithms. M$_{ij}$ is the classifier induced with the feature set $i$ and algorithm $j$, $i=\text{1,...,12}$ and $j=\text{1, 2, 3}$, and $w_{ij}$ is the accuracies of the classifier M$_{ij}$. $\hat{M}_{ij}$ is the predicted class by M$_{ij}$, $\hat{M}_{ij} \in \{-1,1\}$.  Emv=Ensemble majority votes, Ewv=Ensemble weighted votes.}

 \begin{table}
    \centering
    \label{tab:allmodels}
    \setlength{\extrarowheight}{0.2em}
    {\small
    \noindent\begin{tabularx}{\textwidth}{X X X X}   \noalign{\hrule height 1pt}   
    & 1. SVMs & 2. RF  & 3. J48 \\ \noalign{\hrule height 1pt}
1. FS$_1$ & M$_{11}$ & M$_{12}$ & M$_{13}$ \\
2. FS$_2$ & M$_{21}$ & M$_{22}$ & M$_{23}$ \\
3. FS$_6$ & M$_{31}$ & M$_{32}$ & M$_{33}$ \\
4. FS$_7$ & M$_{41}$ & M$_{42}$ & M$_{43}$ \\ \hline
5. FS$_3$ & M$_{51}$ & M$_{52}$ & M$_{53}$ \\
6. FS$_4$ & M$_{61}$ & M$_{62}$ & M$_{63}$ \\
7. FS$_5$ & M$_{71}$ & M$_{72}$ & M$_{72}$ \\
8. SELECT & M$_{81}$ & M$_{82}$ & M$_{83}$ \\
9. Hyb$_{37}$ & M$_{91}$ & M$_{92}$ & M$_{93}$ \\
10. Hyb$_S7$ & M$_{101}$ & M$_{102}$ & M$_{103}$ \\
11. Hyb$_17$ & M$_{111}$ & M$_{112}$ & M$_{113}$ \\
12. Ss$_1$ & M$_{121}$ & M$_{122}$ & M$_{123}$ \\ \hline
 Emv8 & $\sum \hat{M}_{i1}$, $i=\text{5,...,12}$ & $\sum\hat{M}_{i2}$, $i=\text{5,...,12}$ & $\sum\hat{M}_{i3}$, $i=\text{5,...,12}$ \\ 
 Ewv8 & $\sum w_{i1}\hat{M}_{i1}$, $i=\text{5,...,12}$ & $\sum w_{i2}\hat{M}_{i2}$, $i=\text{5,...,12}$& $\sum w_{i3}\hat{M}_{i3}$, $i=\text{5,...,12}$ \\ \hline
Emv24 & \multicolumn{3}{c}{ $\sum \hat{M}_{ij}$, $i=\text{5,...,12}$ and $j=\text{1, 2, 3}$ }\\
Ewv24 & \multicolumn{3}{c}{ $\sum w_{ij}\hat{M}_{ij}$, $i=\text{5,...,12}$ and $j=\text{1, 2, 3}$ } \\
\noalign{\hrule height 1pt}     
  \end{tabularx}
  } 
 \end{table}

 \subsection*{Table 3 -  Centers of accuracy clusters from 24 classification models, per species. Range = Maximum - minimum.}   

\begin{table}
   \label{tab:centermodels}
   \centering
   \par
   \setlength{\extrarowheight}{0.2em}
   {\footnotesize
     \noindent\begin{tabular}{ l c c c c c c} \noalign{\hrule height 1.0pt}
Acronym for species & $C_1$ & $C_2$ & $C_3$ & $C_4$ & $C_5$ & Range\\  \noalign{\hrule height 1.0pt}
bfl & 94 & 83 & - & - & - & 15.0 \\ 
cin & 83 & 79 & 75 & 68 & - & 19.0 \\ 
cbr & 93 & 85 & 79 & - & - & 17.0 \\ 
cel & 92 & 87 & 81 & 75 & - & 20.0 \\ 
aae & 95 & 90 & 80 & - & - & 18.0 \\ 
ame & 85 & 78 & 72 & - & - & 20.0 \\ 
api & 92 & 88 & 82 & 73 & - & 22.0 \\ 
bmo & 84 & 79 & 71 & 57 & - & 31.0 \\ 
dme & 91 & 78 & - & - & - & 22.0 \\ 
tca & 89 & 82 & 76 & - & - & 18.0 \\ 
aca & 93 & 86 & 80 & - & - & 16.0 \\ 
xtr & 97 & 87 & 82 & - & - & 18.0 \\ 
gga & 95 & 90 & 85 & 76 & 68 & 27.0 \\ 
cfa & 91 & 83 & 75 & - & - & 22.0 \\ 
eca & 93 & 86 & 77 & - & - & 20.0 \\ 
mdo & 87 & 79 & 71 & - & - & 21.0 \\ 
mml & 89 & 82 & 75 & - & - & 17.0 \\ 
ggo & 89 & 77 & 66 & - & - & 27.0 \\ 
hsa & 88 & 77 & - & - & - & 16.0 \\ 
ptr & 89 & 82 & 73 & - & - & 23.0 \\ 
oan & 88 & 83 & 77 & 70 & - & 23.0 \\ 
cgr & 92 & 88 & 84 & 78 & - & 16.0 \\ 
mmu & 85 & 79 & 72 & - & - & 17.0 \\ 
rno & 93 & 88 & 81 & - & - & 17.0 \\ 
bta & 84 & 80 & 75 & 68 & - & 18.0 \\ 
oar & 91 & 86 & 77 & - & - & 18.0 \\ 
ssc & 90 & 85 & 79 & 64 & - & 29.0 \\ 
dre & 93 & 86 & 80 & - & - & 17.0 \\ 
ola & 92 & 88 & 80 & 68 & - & 26.0 \\ 
ppt & 93 & 84 & 76 & - & - & 20.0 \\ 
aly & 95 & 88 & 81 & - & - & 17.0 \\ 
ath & 94 & 83 & - & - & - & 15.0 \\ 
mes & 98 & 91 & 85 & - & - & 14.0 \\ 
gma & 91 & 86 & 79 & - & - & 18.0 \\ 
mtr & 86 & 82 & 72 & - & - & 21.0 \\ 
lus & 97 & 84 & - & - & - & 18.0 \\ 
mdm & 98 & 85 & - & - & - & 15.0 \\ 
ppe & 95 & 87 & 80 & - & - & 18.0 \\ 
ptc & 94 & 83 & - & - & - & 16.0 \\ 
stu & 93 & 87 & 82 & - & - & 16.0 \\ 
vvi & 93 & 86 & 78 & - & - & 20.0 \\ 
bdi & 91 & 87 & 75 & - & - & 22.0 \\ 
osa & 87 & 77 & - & - & - & 16.0 \\ 
sbi & 96 & 89 & 81 & - & - & 20.0 \\ 
zma & 96 & 82 & - & - & - & 17.0 \\ \noalign{\hrule height 1.0pt}
     \end{tabular}} 
   \end{table}

 \subsection*{Table 4 - Centers of accuracy clusters obtained from classification models induced with examples from different species, per combination of feature set and learning algorithm. Range = Maximum - minimum.}  
 
\begin{table}
   \label{tab:centerspecies}
   \centering
   \par
   \setlength{\extrarowheight}{0.2em}
   {\footnotesize
     \noindent\begin{tabular}{ l |l |r r r r r r r } \noalign{\hrule height 1.0pt}
FEATURE SET & ALGORITHM & $c_1$ & $c_2$ & $c_3$ & $c_4$ & $c_5$ & $c_6$ & Range \\ \noalign{\hrule height 1.0pt}
FS1  & \multirow{8}{*}{SVM} & 95 & 88 & 78 &   -   &   -   &   -   & 21 \\
FS2  &                                 & 96 & 92 & 87 & 80 &   -   &   -   & 20 \\
FS3  &                                 & 95 & 90 & 85 &   -   &   -   &   -   & 15 \\
FS4  &                                 & 92 & 86 & 81 & 77 &   -   &   -   & 22 \\
FS5  &                                 & 94 & 90 & 86 & 80 &   -   &   -   & 20 \\
FS6  &                                 & 93 & 88 & 83 &   -   &   -   &   -   & 17 \\
FS7  &                                 & 95 & 88 &   - &   -   &   -   &   -   & 16 \\
SELECT &                             & 96 & 92 & 86 & 80 &   -   &   -   & 20 \\ \hline
FS1  &\multirow{8}{*}{RF}     & 97 & 92 & 87 & 82 & 72 &   -   & 30 \\
FS2  &                                 & 97 & 93 & 89 & 83 &   -   &   -   & 20 \\
FS3  &                                 & 95 & 88 & 84 &   -   &   -   &   -   & 18 \\
FS4  &                                 & 91 & 87 & 84 & 79 &   -   &   -   & 18 \\
FS5  &                                 & 92 & 85 & 77 &   -   &   -   &   -   & 19 \\
FS6  &                                 & 95 & 88 &   -   &   -   &   -   &   -   & 16 \\
FS7  &                                 & 96 & 89 &   -   &   -   &   -   &   -   & 14 \\
SELECT&\multirow{8}{*}{J48} & 96 & 92 & 86 & 78 &   -   &   -   & 21 \\ \hline
FS1  &                                 & 98 & 91 & 85 & 75 & 67 & 57 & 41 \\
FS2  &                                 & 96 & 90 & 84 & 77 &   -   &   -   & 24 \\
FS3  &                                 & 97 & 92 & 87 & 81 &   -   &   -   & 21 \\
FS4  &                                 & 84 & 79 & 75 & 69 &   -   &   -   & 21 \\
FS5  &                                 & 83 & 78 & 75 & 71 &   -   &   -   & 17 \\
FS6  &                                 & 97 & 93 & 89 & 83 & 78 & 72 & 27 \\
FS7  &                                 & 96 & 91 & 87 & 81 &   -   &   -   & 21 \\
SELECT &                             & 97 & 92 & 86 & 80 & 74 &   -   & 26 \\ \noalign{\hrule height 1.0pt}
     \end{tabular}} 
   \end{table}

\subsection*{Table 5 - Relative feature importance ($RFI$) distributions. Omitting those of $ RFI\le$ 0.1 for all species.}   

\begin{table}
 \label{tab:RFI}   
 \centering
 \par
 {\scriptsize
 \noindent\begin{tabular}{r|l|r|r|r|r|r|r|r|r}  
 \noalign{\hrule height 1.0pt}
\multirow{2}{*}{N\textsuperscript{\underline{o}}} & \multirow{2}{*}{FEATURE} & \multicolumn{8}{c}{$RFI$ intervals} \\ \cline{3-10}
        & & $\le$0.3 & (0.3,0.4 & (0.4,0.5] & (0.5,0.6] & (0.6,0.7] & (0.7,0.8] & (0.8,0.9] & $\ge$1.0 \\  \hline 
1 & $p$     & 4 & 0 & 4 & 2 & 9 & 13 & 16 & 51 \\ \hline 
2 & $MFEI_{1}$ & 4 & 0 & 4 & 0 & 7 & 4 & 22 & 58 \\ \hline 
3 & $dP$ & 33 & 29 & 20 & 11 & 7 & 0 & 0 & 0 \\ \hline 
4 & $ZG$ & 36 & 9 & 9 & 16 & 7 & 2 & 4 & 18 \\ \hline 
5 & $orf$ & 36 & 31 & 4 & 11 & 9 & 2 & 4 & 2 \\ \hline 
6 & $dG$ & 47 & 20 & 13 & 11 & 7 & 2 & 0 & 0 \\ \hline 
7 & $ZP$ & 53 & 13 & 9 & 2 & 7 & 2 & 0 & 13 \\ \hline 
8 & $Avg$\_Bp\_Stem & 53 & 33 & 9 & 4 & 0 & 0 & 0 & 0 \\ \hline 
9 & $EAFE$ & 64 & 20 & 4 & 7 & 2 & 2 & 0 & 0 \\ \hline 
10 & $MFEI_{3}$ & 67 & 13 & 9 & 4 & 2 & 0 & 2 & 2 \\ \hline 
11 & $MFEI_{4}$ & 80 & 9 & 7 & 0 & 0 & 2 & 2 & 0 \\ \hline 
12 & $|A-U|/L$ & 80 & 16 & 4 & 0 & 0 & 0 & 0 & 0 \\ \hline
13 & $ZQ$ & 82 & 2 & 4 & 2 & 0 & 2 & 2 & 4 \\ \hline 
14 & $ZD$ & 84 & 0 & 2 & 4 & 2 & 2 & 2 & 2 \\ \hline 
15 & $A_{(((}$ & 89 & 9 & 0 & 2 & 0 & 0 & 0 & 0 \\ \hline 
16 & $Tm$ & 93 & 7 & 0 & 0 & 0 & 0 & 0 & 0 \\ \hline 
17 & $dQ$ & 93 & 7 & 0 & 0 & 0 & 0 & 0 & 0 \\ \hline 
18 & $Diversity$ & 98 & 2 & 0 & 0 & 0 & 0 & 0 & 0 \\ \hline 
19 & $U_{(((}$ & 98 & 0 & 2 & 0 & 0 & 0 & 0 & 0 \\ \hline 
20 & $MFE$ & 98 & 0 & 2 & 0 & 0 & 0 & 0 & 0 \\ \hline 
21 & $\%(A-U)/stems$ & 100 & 0 & 0 & 0 & 0 & 0 & 0 & 0 \\ \hline 
22 & $dD$ & 100 & 0 & 0 & 0 & 0 & 0 & 0 & 0 \\ \hline 
23 & $Diff$ & 100 & 0 & 0 & 0 & 0 & 0 & 0 & 0 \\ \hline 
24 & $Tm/L$ & 100 & 0 & 0 & 0 & 0 & 0 & 0 & 0 \\ \hline 
25 & $\%UU$ & 100 & 0 & 0 & 0 & 0 & 0 & 0 & 0 \\ \hline 
26 & $dm$ & 100 & 0 & 0 & 0 & 0 & 0 & 0 & 0 \\ \hline 
27 & $\%G+C$ & 100 & 0 & 0 & 0 & 0 & 0 & 0 & 0 \\ \hline 
28 & $MFEI_{4}$ & 100 & 0 & 0 & 0 & 0 & 0 & 0 & 0 \\ \hline 
29 & $A_{...}$ & 100 & 0 & 0 & 0 & 0 & 0 & 0 & 0 \\ \hline 
30 & $\%GA$ & 100 & 0 & 0 & 0 & 0 & 0 & 0 & 0 \\ \hline 
31 & $C_{...}$ & 100 & 0 & 0 & 0 & 0 & 0 & 0 & 0 \\ \hline 
32 & $dH/L$ & 100 & 0 & 0 & 0 & 0 & 0 & 0 & 0 \\ \hline 
33 & $G_{...}$ & 100 & 0 & 0 & 0 & 0 & 0 & 0 & 0 \\ \hline 
34 & $\%UA$ & 100 & 0 & 0 & 0 & 0 & 0 & 0 & 0 \\ \hline \noalign{\hrule height 1.0pt}
\end{tabular}
 } 
\end{table}


\section*{Additional Files}
\subsection*{Additional file 1 --- Pre-miRNAs description} 
\subsection*{Additional file 2 --- Pseudo description} 


\begin{table}
  \small
  \caption{Phylum/division, subphylum/class, species, acronyms, number of positive examples available at miRBase 20, mean and standard deviation of the length distributions. NR=Non-Redundant.}
 \label{tab:characteristics}
  {\small
\begin{tabularx}{\textwidth}{X|X|l|l|r|r|r} \noalign{\hrule height 1.0pt}   
 \multirow{2}{*}{Phylum/Division}& \multirow{2}{*}{Subphylum/Class} & \multirow{2}{*}{Species genus} & \multirow{2}{*}{Acronym} & \multicolumn{2}{c|}{ \#pre-miRNA }   & \multicolumn{1}{c}{Length} \\ \cline{5-6}
              &                 &         &           &  All  & NR          &  (Mean $\pm$ SD) \\ \noalign{\hrule height 1.0pt}
\multirow{29}{*}{Chordate}& Cephalochordata& Branchiostoma floridae   & bfl   & 156 & 143 & 87 $\pm$ 13 \\\cline{2-7}
& Urochordata      & Ciona intestinalis  & cin   & 346 & 331 & 63  $\pm$ 16 \\ \cline{2-7}
&\multirow{2}{*}{Nematoda} & Caenorhabditis briggsae & cbr   & 177  & 148 & 92 $\pm$ 19 \\ 
& & Caenorhabditis elegans    & cel   & 233 & 214 & 89 $\pm$ 17 \\ \cline{2-7}
&\multirow{6}{*}{Hexapoda}& Aedes egypti  & aae  & 101 & 90 & 94 $\pm$ 21 \\
& & Apis mellifera                & ame  & 218 & 215 & 100 $\pm$ 20 \\
&  & Acyrthosiphon pisum     & api   & 117 & 101 & 66 $\pm$ 9 \\
&  & Bombyx mori               & bmo & 489 & 432 & 100 $\pm$ 22 \\
&& Drosophila melanogaster & dme     &  238 & 236 & 95 $\pm$ 23 \\
&& Tribolium castaneum     & tca     & 220 & 210 & 95 $\pm$ 22 \\ \cline{2-7}
&\multirow{19}{*}{Vertebrate}&Anolis carolinensis  & aca & 282 & 272 & 89 $\pm$ 9 \\
&& Xenopus tropicalis      & xtr     & 189 & 163 & 83 $\pm$ 11 \\
&& Gallus gallus           & gga     & 734 & 695  & 92 $\pm$ 17 \\
&& Canis familiaris        & cfa     & 324 & 280 & 69 $\pm$ 14 \\
&& Equus caballus          & eca     & 341 & 298 & 78 $\pm$ 15 \\
&& Monodelphis domestica   & mdo   & 460 & 370  & 67 $\pm$ 12 \\
&& Macaca mulatta          & mml     &  615 & 524 & 86 $\pm$ 17 \\
&& Gorilla gorilla         & ggo     &  332 & 313 & 105 $\pm$ 12 \\
&& Homo sapiens            & hsa   & 1,872  & 1,640 & 82 $\pm$ 17 \\
&& Pan troglodytes         & ptr     & 659 & 542 & 90 $\pm$ 17 \\
&& Ornithorhynchus anatinus& oan     & 396  & 327 & 100 $\pm$ 24\\
&& Cricetulus griseus      & cgr     &  200 & 199 & 82 $\pm$ 12 \\
&& Mus musculus            & mmu     & 1,186 & 1,078  & 83 $\pm$ 19 \\
&& Rattus norvegicus       & rno     & 449 & 428  & 92 $\pm$ 17 \\
&& Bos taurus              & bta     & 798 & 710  & 80 $\pm$ 13 \\
&& Ovis aries              & oar     & 105 & 96  & 97 $\pm$ 18 \\
&& Sus scrofa              & ssc     &  280 & 247  & 81 $\pm$ 10 \\
&& Danio rerio             & dre     &  346  & 240 & 93 $\pm$ 18 \\
&& Oryzias latipes         & ola     &  168  & 146 & 95 $\pm$ 9 \\\hline
Bryophyta& Musci & Physcomitrella patens & ppt & 229 & 204 & 161 $\pm$ 56 \\ \hline
\multirow{15}{*}{Angiospermae}&\multirow{11}{*}{Eudicotyledons}&Arabidopsis lyrata &aly& 298 & 177 & 183 $\pm$ 100 \\
&& Arabidopsis thaliana    & ath     & 298 & 257 & 183 $\pm$ 103 \\
&& Manihot esculenta       & mes     & 153 & 109  & 117 $\pm$ 38 \\
&& Glycine max             & gma     &  505 & 361 & 131 $\pm$ 47 \\
&& Medicago truncatula     & mtr     & 672 & 373 & 165 $\pm$ 91 \\
&& Linum usitatissimum     & lus     &  124 & 100  & 144 $\pm$ 34 \\
&& Malus domestica         & mdm   &  206 & 90  & 130 $\pm$ 66 \\
&& Prunus persica          & ppe     &  180 & 147  & 136 $\pm$ 51 \\
&& Populus trichocarpa     & ptc     & 352 & 246 & 128 $\pm$ 46 \\
&& Solanum tuberosum       & stu     &  224 & 163 & 95 $\pm$ 43 \\
&& Vitis vinifera          & vvi     & 163 & 131 & 127 $\pm$ 56 \\
&\multirow{4}{*}{Monocotyledons} &Brachypodium distachyon & bdi &  258 & 228  & 178 $\pm$ 101 \\ \cline{2-7}
&& Oryza sativa            & osa     & 592  & 482 & 153 $\pm$ 77 \\
&& Sorghum bicolor         & sbi     & 205 & 174 & 142 $\pm$ 54 \\
&& Zea mays                & zma     & 172  & 133  & 132 $\pm$ 45 \\ \noalign{\hrule height 1.0pt}
  \end{tabularx}  } 
  \end{table}


\begin{table}
  \small
  \caption{Phylum/division, subphylum/class, species, acronyms, number of redundant negative examples out of 1,000 sequences excised from CDS or pseudo genes and the corresponding website link for download.}
  \label{tab:charpseudo}
  {\small \begin{tabularx}{\textwidth}{ X | X | l | l | r | c} \noalign{\hrule height 1.0pt}   
Phylum/Division & Subphylum/Class & Species genus & Acronym & \#Redundant & Web Source \\ \noalign{\hrule height 1.0pt}  
\multirow{29}{*}{Chordate}& Cephalochordata& Branchiostoma floridae&bfl&1&\href{ftp://ftp.jgi-psf.org/pub/compgen/metazome/v3.0/Bfloridae/annotation/Bfloridae_17_cds.fa.gz}{Metazome v3.0} \\\cline{2-6}
& Urochordata&Ciona intestinalis&cin&3&\href{ftp://ftp.jgi-psf.org/pub/compgen/metazome/v3.0/Cintestinalis/annotation/Cintestinalis_45_cds.fa.gz}{Metazome v3.0} \\ \cline{2-6}
&\multirow{2}{*}{Nematoda}&Caenorhabditis briggsae&cbr&7&\href{ftp://ftp.jgi-psf.org/pub/compgen/metazome/v3.0/Cbriggsae/annotation/Cbriggsae_13_cds.fa.gz}{Metazome v3.0} \\ 
& & Caenorhabditis elegans&cel&1&\href{ftp://ftp.jgi-psf.org/pub/compgen/metazome/v3.0/Celegans/annotation/Celegans_12_cds.fa.gz}{Metazome v3.0} \\ \cline{2-6}
&\multirow{6}{*}{Hexapoda}&Aedes egypti&aae&4&\href{ftp://ftp.jgi-psf.org/pub/compgen/metazome/v3.0/Aaegypti/annotation/Aaegypti_4_cds.fa.gz}{Metazome v3.0} \\
& & Apis mellifera&ame&395&\href{ftp://ftp.ncbi.nlm.nih.gov/genomes/Apis_mellifera/other/pseudo_without_product.fa.gz}{NCBI} \\
&  & Acyrthosiphon pisum&api&43&\href{ftp://ftp.ncbi.nlm.nih.gov/genomes/Acyrthosiphon_pisum/other/pseudo_without_product.fa.gz}{NCBI} \\
&  & Bombyx mori&bmo&5&\href{ftp://ftp.jgi-psf.org/pub/compgen/metazome/v3.0/Bmori/annotation/Bmori_18_cds.fa.gz}{Metazome v3.0} \\
&& Drosophila melanogaster&dme&3&\href{ftp://ftp.jgi-psf.org/pub/compgen/metazome/v3.0/Dmelanogaster/annotation/Dmelanogaster_3_cds.fa.gz}{Metazome v3.0} \\
&& Tribolium castaneum&tca&0&\href{ftp://ftp.jgi-psf.org/pub/compgen/metazome/v3.0/Tcastaneum/annotation/Tcastaneum_22_cds.fa.gz}{Metazome v3.0} \\ \cline{2-6}
&\multirow{19}{*}{Vertebrate}&Anolis carolinensis&aca&35&\href{ftp://ftp.ncbi.nlm.nih.gov/genomes/Anolis_carolinensis/other/pseudo_without_product.fa.gz}{NCBI} \\
&& Xenopus tropicalis&xtr&2&\href{ftp://ftp.jgi-psf.org/pub/compgen/metazome/v3.0/Xtropicalis/annotation/Xtropicalis_14_cds.fa.gz}{Metazome v3.0} \\
&& Gallus gallus&gga&3&\href{ftp://ftp.jgi-psf.org/pub/compgen/metazome/v3.0/Ggallus/annotation/Ggallus_25_cds.fa.gz}{Metazome v3.0} \\
&& Canis familiaris&cfa&4&\href{ftp://ftp.jgi-psf.org/pub/compgen/metazome/v3.0/Cfamiliaris/annotation/Cfamiliaris_23_cds.fa.gz}{Metazome v3.0} \\
&& Equus caballus&eca&14&\href{ftp://ftp.ncbi.nlm.nih.gov/genomes/Equus_caballus/other/pseudo_without_product.fa.gz}{NCBI} \\
&& Monodelphis domestica&mdo&8& \href{ftp://ftp.jgi-psf.org/pub/compgen/metazome/v3.0/Mdomestica/annotation/Mdomestica_7_cds.fa.gz}{Metazome v3.0}  \\
&& Macaca mulatta&mml&80&\href{ftp://ftp.ncbi.nlm.nih.gov/genomes/Macaca_mulatta/other/pseudo_without_product.fa.gz}{NCBI} \\
&& Gorilla gorilla&ggo&61&\href{ftp://ftp.ncbi.nlm.nih.gov/genomes/Gorilla_gorilla/other/pseudo_without_product.fa.gz}{NCBI} \\
&& Homo sapiens&hsa&3&\href{ftp://ftp.jgi-psf.org/pub/compgen/metazome/v3.0/Hsapiens/annotation/Hsapiens_11_cds.fa.gz}{Metazome v3.0}  \\
&& Pan troglodytes&ptr&88&\href{ftp://ftp.ncbi.nlm.nih.gov/genomes/Pan_troglodytes/other/pseudo_without_product.fa.gz}{NCBI}\\
&& Ornithorhynchus anatinus&oan&50&\href{ftp://ftp.ncbi.nlm.nih.gov/genomes/Ornithorhynchus_anatinus/other/pseudo_without_product.fa.gz}{NCBI} \\
&& Cricetulus griseus&cgr&12&\href{ftp://ftp.ncbi.nlm.nih.gov/genomes/Cricetulus_griseus/other/pseudo_without_product.fa.gz}{NCBI} \\
&& Mus musculus&mmu&5&\href{ftp://ftp.jgi-psf.org/pub/compgen/metazome/v3.0/Mmusculus/annotation/Mmusculus_6_cds.fa.gz}{Metazome v3.0} \\
&& Rattus norvegicus&rno&3&\href{ftp://ftp.jgi-psf.org/pub/compgen/metazome/v3.0/Rnorvegicus/annotation/Rnorvegicus_10_cds.fa.gz}{Metazome v3.0} \\
&& Bos taurus&bta&94&\href{ftp://ftp.ncbi.nlm.nih.gov/genomes/Bos_taurus/other/pseudo_without_product.fa.gz}{NCBI} \\
&& Ovis aries&oar&70&\href{ftp://ftp.ncbi.nlm.nih.gov/genomes/Ovis_aries/other/pseudo_without_product.fa.gz}{NCBI} \\
&& Sus scrofa&ssc&62&\href{ftp://ftp.ncbi.nlm.nih.gov/genomes/Sus_scrofa/other/pseudo_without_product.fa.gz}{NCBI} \\
&& Danio rerio&dre&3&\href{ftp://ftp.jgi-psf.org/pub/compgen/metazome/v3.0/Drerio/annotation/Drerio_9_cds.fa.gz}{Metazome v3.0}  \\
&& Oryzias latipes&ola&0&\href{ftp://ftp.jgi-psf.org/pub/compgen/metazome/v3.0/Olatipes/annotation/Olatipes_8_cds.fa.gz}{Metazome v3.0}  \\\hline
Bryophyta&Musci&Physcomitrella patens&ppt&3&\href{ftp://ftp.jgi-psf.org/pub/compgen/phytozome/v9.0/Ppatens/annotation/Ppatens_152_cds.fa.gz}{Phytozome v9.0} \\ \hline
\multirow{15}{*}{Angiospermae}&\multirow{11}{*}{Eudicotyledons}&Arabidopsis lyrata &aly&5&\href{ftp://ftp.jgi-psf.org/pub/compgen/phytozome/v9.0/Alyrata/annotation/Alyrata_107_cds.fa.gz}{Phytozome v9.0} \\
&& Arabidopsis thaliana&ath&4&\href{ftp://ftp.jgi-psf.org/pub/compgen/phytozome/v9.0/Athaliana/annotation/Athaliana_167_cds.fa.gz}{Phytozome v9.0} \\
&& Manihot esculenta&mes&3&\href{ftp://ftp.jgi-psf.org/pub/compgen/phytozome/v9.0/Mesculenta/annotation/Mesculenta_147_cds.fa.gz}{Phytozome v9.0} \\
&& Glycine max&gma&4&\href{ftp://ftp.jgi-psf.org/pub/compgen/phytozome/v9.0/Gmax/annotation/Gmax_189_cds.fa.gz}{Phytozome v9.0} \\
&& Medicago truncatula&mtr&2&\href{ftp://ftp.jgi-psf.org/pub/compgen/phytozome/v9.0/Mtruncatula/annotation/Mtruncatula_198_cds.fa.gz}{Phytozome v9.0}  \\
&& Linum usitatissimum&lus&2&\href{ftp://ftp.jgi-psf.org/pub/compgen/phytozome/v9.0/Lusitatissimum/annotation/Lusitatissimum_200_cds.fa.gz}{Phytozome v9.0} \\
&& Malus domestica&mdm&3&\href{ftp://ftp.jgi-psf.org/pub/compgen/phytozome/v9.0/Mdomestica/annotation/Mdomestica_196_cds.fa.gz}{Phytozome v9.0}  \\
&& Prunus persica&ppe&15&\href{ftp://ftp.jgi-psf.org/pub/compgen/phytozome/v9.0/Ppersica/annotation/Ppersica_139_cds.fa.gz}{Phytozome v9.0}  \\
&& Populus trichocarpa&ptc&3&\href{ftp://ftp.jgi-psf.org/pub/compgen/phytozome/v9.0/Ptrichocarpa/annotation/Ptrichocarpa_210_cds.fa.gz}{Phytozome v9.0} \\
&& Solanum tuberosum&stu&2&\href{ftp://ftp.jgi-psf.org/pub/compgen/phytozome/v9.0/Stuberosum/annotation/Stuberosum_206_cds.fa.gz}{Phytozome v9.0} \\
&&Vitis vinifera&vvi&0&\href{ftp://ftp.jgi-psf.org/pub/compgen/phytozome/v9.0/Vvinifera/annotation/Vvinifera_145_cds.fa.gz}{Phytozome v9.0} \\
&\multirow{4}{*}{Monocotyledons}&Brachypodium distachyon&bdi&2&\href{ftp://ftp.jgi-psf.org/pub/compgen/phytozome/v9.0/Bdistachyon/annotation/Bdistachyon_192_cds.fa.gz}{Phytozome v9.0} \\ \cline{2-6}
&& Oryza sativa&osa&4&\href{ftp://ftp.jgi-psf.org/pub/compgen/phytozome/v9.0/Osativa/annotation/Osativa_204_cds.fa.gz}{Phytozome v9.0} \\
&& Sorghum bicolor&sbi&3&\href{ftp://ftp.jgi-psf.org/pub/compgen/phytozome/v9.0/Sbicolor/annotation/Sbicolor_79_cds.fa.gz}{Phytozome v9.0} \\
&&Zea mays&zma&1&\href{ftp://ftp.jgi-psf.org/pub/compgen/phytozome/v9.0/Zmays/annotation/Zmays_181_cds.fa.gz}{Phytozome v9.0} \\\noalign{\hrule height 1.0pt}
  \end{tabularx}  } 
  \end{table}



\end{bmcformat}

\end{document}